\documentclass{emulateapj}
\usepackage{apjfonts}
\usepackage{amsmath}
\usepackage{booktabs}
\usepackage{color}
\usepackage{graphicx}
\usepackage{amsmath}
\begin{document}

%% LaTeX will automatically break titles if they run longer than
%% one line. However, you may use \\ to force a line break if
%% you desire.

\title{Compressible Relativistic Magnetohydrodynamic Turbulence in Magnetically-Dominated Plasmas 
And Implications for 
%A New Regime}       
A Strong-Coupling Regime
}       
%% Use \author, \affil, and the \and command to format
%% author and affiliation information.
%% Note that \email has replaced the old \authoremail command
%% from AASTeX v4.0. You can use \email to mark an email address
%% anywhere in the paper, not just in the front matter.
%% As in the title, use \\ to force line breaks.

\author{Makoto Takamoto}
\affil{Department of Earth and Planetary Science, The University of Tokyo, Hongo, Bunkyo-ku, Tokyo, 113-0033, Japan}
\email{mtakamoto@eps.s.u-tokyo.ac.jp}

\author{Alexandre Lazarian}
\affil{Department of Astronomy, University of Wisconsin, 475 North Charter Street, Madison, WI 53706, USA}
\email{alazarian@facstaff.wisc.edu}

%% Notice that each of these authors has alternate affiliations, which
%% are identified by the \altaffilmark after each name.  Specify alternate
%% affiliation information with \altaffiltext, with one command per each
%% affiliation.

%\altaffiltext{1}{Visiting Astronomer, Cerro Tololo Inter-American Observatory.
%CTIO is operated by AURA, Inc.\ under contract to the National Science
%Foundation.}
%\altaffiltext{2}{Society of Fellows, Harvard University.}
%\altaffiltext{3}{present address: Center for Astrophysics,
%    60 Garden Street, Cambridge, MA 02138}
%\altaffiltext{4}{Visiting Programmer, Space Telescope Science Institute}
%\altaffiltext{5}{Patron, Alonso's Bar and Grill}

%% Mark off your abstract in the ``abstract'' environment. In the manuscript
%% style, abstract will output a Received/Accepted line after the
%% title and affiliation information. No date will appear since the author
%% does not have this information. The dates will be filled in by the
%% editorial office after submission.

\begin{abstract}
In this Letter, 
%we report compressible mode effects on relativistic magnetohydrodynamic (RMHD) turbulence in high-$\sigma$ plasmas 
we report compressible mode effects on relativistic magnetohydrodynamic (RMHD) turbulence in Poynting-dominated plasmas 
using 3-dimensional numerical simulations. 
We decomposed fluctuations in the turbulence into 3 MHD modes (fast, slow, and Alfv\'en) 
%\textcolor{red}{%
%assuming the weak-coupling between each modes, 
following the procedure mode decomposition in \citep{2002PhRvL..88x5001C}, 
%}%
and analyzed their energy spectra and structure functions separately. 
%We also analyzed the ratio of \textcolor{red}{compressible mode} to Alfv\'en mode energy with respect to \textcolor{}Alfv\'en Mach number of turbulence. 
We also analyzed the ratio of compressible mode to Alfv\'en mode energy with respect to its Mach number. 
%We found the ratio of compressible mode increases not only with the Alfv\'en Mach number but with the background $\sigma$-parameter, 
We found the ratio of compressible mode increases not only with the Alfv\'en Mach number but with the background magnetization, 
which indicates a strong coupling between the fast and Alfv\'en modes 
and appearance of a new regime of RMHD turbulence in Poynting-dominated plasmas 
where the fast and Alfv\'en modes strongly couples and cannot be separated, different from the non-relativistic MHD case.
%\textcolor{red}{%
This finding will affect particle acceleration efficiency obtained by assuming Alfv\'enic critical balance turbulence, 
and will change the resulting photon spectra by non-thermal electrons. 
%}%
\end{abstract}

%% Keywords should appear after the \end{abstract} command. The uncommented
%% example has been keyed in ApJ style. See the instructions to authors
%% for the journal to which you are submitting your paper to determine
%% what keyword punctuation is appropriate.

\keywords{magnetic fields, magnetohydrodynamics (MHD), relativistic processes, plasmas, turbulence}

%% From the front matter, we move on to the body of the paper.
%% In the first two sections, notice the use of the natbib \citepp
%% and \citept commands to identify citations.  The citations are
%% tied to the reference list via symbolic KEYs. The KEY corresponds
%% to the KEY in the \bibitem in the reference list below. We have
%% chosen the first three characters of the first author's name plus
%% the last two numeral of the year of publication as our KEY for
%% each reference.

%% Authors who wish to have the most important objects in their paper
%% linked in the electronic edition to a data center may do so by tagging
%% their objects with \objectname{} or \object{}.  Each macro takes the
%% object name as its required argument. The optional, square-bracket 
%% argument should be used in cases where the data center identification
%% differs from what is to be printed in the paper.  The text appearing 
%% in curly braces is what will appear in print in the published paper. 
%% If the object name is recognized by the data centers, it will be linked
%% in the electronic edition to the object data available at the data centers  
%%
%% Note that for sources with brackets in their names, e.g. [WEG2004] 14h-090,
%% the brackets must be escaped with backslashes when used in the first
%% square-bracket argument, for instance, \object[\[WEG2004\] 14h-090]{90}).
%%  Otherwise, LaTeX will issue an error. 

\section{\label{sec:sec1}Introduction}
Turbulence is ubiquitous in many plasma phenomena. 
Over the past few decades, a considerable number of studies have been conducted on the properties of non-relativistic magnetohydrodynamic (MHD) turbulence, 
and they have revealed many interesting and essential properties, 
for example, the critical-balance condition of strong turbulence \citep{1995ApJ...438..763G,2010ApJ...722L.110B,2014ApJ...784L..20B}, 
the existence of weak turbulence \citep{1964SvA.....7..566I,1965PhFl....8.1385K,2000JPlPh..63..447G,2016PhRvL.116j5002M}, 
and mode coupling between Alfv\'enic and compressible modes \citep{2002PhRvL..88x5001C,2003MNRAS.345..325C}. 
%In particular, 
%many astrophysical phenomena are considered to be in a turbulent state 
%because of their large scale size comparing with typical plasma scales. 
%On the other hand, 
The concern with relativistic turbulence has recently been growing 
because of the development of high-power laser plasma devices and observation devices for high-energy astrophysical phenomena. 
In particular, 
the observation of high energy astrophysical phenomena, such as gamma-ray bursts and flares of relativistic jets, indicates that 
there will be strong turbulence 
that is necessary for acceleration of electrons emitting non-thermal photons observed from those phenomena \citep{2012ApJ...754..114H,2015ApJ...808L..18A}. 
Recent theoretical studies of those phenomena indicate that 
the background plasma of those phenomena is Poynting-energy dominated \citep{1984ApJ...283..694K,1984ApJ...283..710K,2015ApJ...803...30K}
in which the relativistic magnetization parameter, $\sigma \equiv B_0^2/4 \pi \rho h c^2 \gamma^2$, is larger than unity 
where $B_0$ is the background magnetic field, $\rho$ is the rest mass density, $h$ is the specific enthalpy, 
$c$ is the velocity of light, and $\gamma$ is the Lorentz factor
\footnote{
%\textcolor{red}{%
Note that the $\sigma$-parameter is originally defined as a ratio between Poynting-flux to particle energy flux. 
The form introduced in this Letter is a reduced one assuming MHD approximation ${\bf E} = {\bf v}/c \times {\bf B}$ 
that can be used in a static background flow. 
It also becomes the square of relativistic 4-Alfv\'en velocity in the fluid comoving frame, 
and is popular in high energy astrophysics community. 
%}%
}.
Although there are several works investigating turbulence in relativistic MHD plasma \citep{2011ApJ...734...77I,2012ApJ...744...32Z,2013ApJ...763L..12Z,2013ApJ...766L..10R,2015ApJ...815...16T} 
and force-free plasma \citep{1998PhRvD..57.3219T,2005ApJ...621..324C,2014ApJ...780...30C}, 
many properties of the relativistic MHD turbulence and the dependence on the background $\sigma$-parameter are still unclear, 
%\textcolor{red}{%
in particular, in the case of Poynting-dominated RMHD turbulence. 
%}%
%Furthermore, 
%there are few work investigating mode coupling between Alfv\'enic and compressible modes in relativistic MHD turbulence \citep{2015ApJ...815...16T} 
%that is important to determining the non-thermal electron energy spectrum accelerated by the Fermi 2nd-order process 
%\citep{1949PhRv...75.1169F,1954ApJ...119....1F,2004ApJ...610..550P}. 
%\textcolor{blue}{%
%Furthermore, there is work studying mode coupling between Alfv\'enic and compressible modes in relativistic MHD turbulence \citep{2015ApJ...815...16T} 
%done in the framework of turbulent reconnection which has implications e.g. for the non-thermal electron energy spectrum accelerated by the Fermi 2nd-order process 
%\citep{1949PhRv...75.1169F,1954ApJ...119....1F,2004ApJ...610..550P}. 
%}%

In this Letter, 
we report our findings on RMHD turbulence, in particular, the properties of each MHD characteristic mode 
in low-$\beta$ plasma but covering low-$\sigma$ to high-$\sigma$ plasma for the first time. 
We performed a series of numerical relativistic MHD simulations of turbulence, 
and analyzed the spectral properties of each mode, their structure function, and eddy-scale in terms of local mean field. 
%\textcolor{red}{%
We also discuss a possibility of a strong coupling between the fast and Alfv\'en modes 
and appearance of a new regime of RMHD turbulence in Poynting-dominated plasmas 
where the fast and Alfv\'en modes strongly couples and cannot be separated, different from the non-relativistic MHD case. 
%}%

\section{\label{sec:sec2}Numerical Setup}
The plasma is modeled by the ideal RMHD approximation with the TM equation of state \citep{2005ApJS..160..199M} 
that allows us to simulate the relativistic perfect gas equation of state \citep{synge1957relativistic} with less than 4 \% error. 
%%\textcolor{red}{%
%The RMHD equations are given as: 
%\begin{align}
%  &\partial_t (\rho \gamma) + \nabla \cdot [\rho \gamma {\bf v}] = 0, 
%  \label{eq:1.1}                                                               
%  \\
%  &\partial_{\mu} \left[(\rho h + b^2) u^{\mu} u^{\nu} - b^{\mu} b^{\nu} + \left(p_g + \frac{b^2}{2} \right) \eta^{\mu \nu} \right] = 0, 
%  \label{eq:1.2}
%  \\
%  &\partial_t {\bf B} + \nabla \times ({\bf v} \times {\bf B}) = 0, 
%  \label{eq:1.3}
%\end{align}
%where $\mu, \nu$ run from 0 to 3 following the Einstein rule, 
%$\eta^{\mu \nu}$ is the flat-space metric, 
%$u^{\mu} = \gamma (1, {\bf v})$ is the four velocity, 
%and $b^{\mu} = \gamma [ {\bf v \cdot B}, {\bf B}/\gamma^2 + ({\bf v \cdot B}){\bf v}]$ is the covariant magnetic field 
%which is the magnetic field in the fluid comoving frame. 
%%}%
The equations are updated by the relativistic HLLD method \citep{2009MNRAS.393.1141M}
in a conservative fashion using the constrained transport algorithm \citep{1988ApJ...332..659E,2005JCoPh.205..509G}. 
The initial background plasma is assumed to be uniform with a uniform magnetic field ${\bf B}_0$, density $\rho_0$, and temperature $k_{\rm B} T = 0.1 m c^2$ 
where $k_B$ is the Boltzmann constant, $T$ is the temperature, $m$ is the particle mass. 
We injected an isotropic turbulent mode component at the initial time-step, so-called \textit{decaying turbulence}. 
Similar to \citep{2015ApJ...815...16T}, 
we injected the turbulence at large scales, $l_{\rm inject} = L/2, L/3, L/4$, where $L$ is the numerical box size, 
and the energy spectrum is assumed to be flat. 
Following \citep{2002PhRvL..88x5001C,2003MNRAS.345..325C}, 
the turbulence is injected only with an Alfv\'en mode velocity component 
that is obtained by the method explained in the next section. 
We consider a cubic numerical domain that is divided by uniform meshes 
%\textcolor{red}{%
whose size is typically $\Delta = L/512$ to obtain fast to Alfv\'en mode power ratio (Figure 1); 
we use higher resolutions, $\Delta = L/1024, L/2048$, to obtain the energy spectra and the structure functions 
provided in Figure 2 
($L/1024$ for $\sigma = 0.2, 1$ and $L/2048$ for $\sigma=3$). 
%}%
The resolution is chosen as sufficient for resolving turbulent-eddies 
responsible for the mode exchange in our simulation 
\footnote{
%\textcolor{red}{%
``\textit{ideal}'' RMHD means that no explicit dissipation processes are included, 
such as the viscosity, thermal conductivity, and resistivity. 
However, the explicit differential scheme we employed here always include 
the numerical grid-scale dissipation, which allows the dissipation at the smallest-eddy 
and direct cascade of energy into smaller scale. 
%}%
}.

\section{Mode decomposition of RMHD Turbulence}
Following \citep{2002PhRvL..88x5001C,2003MNRAS.345..325C}, 
we consider the displacement vectors of slow and fast modes. 
Since there is no average velocity in the background flow, 
the displacement vectors reduce to: 
\begin{align}
  \hat{\xi}_{\rm slow} &\propto k_{||} \hat{\bf k}_{||} + \left[ \frac{u_{\rm slow}^2}{c_{\rm s}^2} \left( \frac{k}{k_{||}} \right)^2 - 1 \right] 
                      \left[ \frac{k_{||}}{k_{\perp}} \right]^2 k_{\perp} \hat{\bf k}_{\perp}
  ,                       
  \label{eq:3.1}
  \\                    
  \hat{\xi}_{\rm fast} &\propto \left[ u_{\rm fast}^2 k^2 (1 + \sigma) - c_{\rm s}^2 k_{\perp}^2 - k^2 \sigma \right] 
                      \left[ \frac{1}{c_{\rm s} k_{||}} \right]^2 k_{||} \hat{\bf k}_{||} + k_{\perp} \hat{\bf k}_{\perp}
  ,                      
  \label{eq:3.2}
\end{align}
where $c_s$ is the relativistic sound velocity, 
$k$ is the absolute value of the wave vector ${\bf k}$, 
and $u_{\rm fast/slow}$ are the characteristic velocities of fast and slow modes \citep{1990rfmf.book.....A}. 
$k_{||}, k_{\perp}$ are taken as the parallel and perpendicular to the background magnetic field direction (${\bf B}_0$). 
Note that Equations (\ref{eq:3.1}) and (\ref{eq:3.2}) reduce to the Equations (1) and (2) of \citep{2002PhRvL..88x5001C} 
if we take the non-relativistic limit and set $c_{\rm s} = 0.1$ and $c_{\rm A} = 1$ where $c_A$ is the Alfv\'en velocity.  
Equations (\ref{eq:3.1}) and (\ref{eq:3.2}) allow us to obtain fast and slow mode velocity components 
by projecting the velocity field on the displacement vectors $\hat{\xi}_{\rm fast/slow}$. 
Similar to the non-relativistic case, 
the displacement vector of the Alfv\'en mode velocity component is given as: $\hat{\xi}_{\rm A} = \hat{\bf k}_{||} \times \hat{\bf k}_{\perp}$ \citep{2001ApJ...554.1175M}. 
%\textcolor{red}{%
In the following, we write the velocity projection onto $\xi_{\rm \{A, fast, slow\}}$ as $\delta v_{\rm \{A, f, s\}}$. 
%}%
Since the form of the linearized mass conservation law and induction equation are the same as the non-relativistic ones, 
the Fourier components of density and magnetic field can be obtained by exactly the same procedure in the non-relativistic case 
given in \citep{2002PhRvL..88x5001C,2003MNRAS.345..325C}. 
%\textcolor{red}{%
%Note that this kind of the mode decomposition is not always possible. 
%However, it can be \textit{statistically} valid 
%if the mode couplings though non-linear terms are weak, such as the non-relativistic MHD turbulence \citep{2002PhRvL..88x5001C}. 
%In the following, we discuss the validity of the mode decomposition by studying the each mode couplings. 
We should stress that, for turbulence, the mode decomposition is valid in a statistical sense as it discussed in \citep{2002PhRvL..88x5001C}.
Moreover, the meaning of the decomposition is different for the case of weak and strong coupling of modes.
When the transfer of energy between the modes is weak as it is the case of non-relativistic turbulence,
the mode decomposition reveals the distinct cascades among Alfv\'en, fast, and slow mode \citep{2002PhRvL..88x5001C}.
In the case of relativistic turbulence, as the coupling increases the decomposition reveals the transfer of energy between the cascades.
%}%

When we calculated the kinetic power, the energy spectrum, and the 2nd-order structure functions of each mode, 
we projected the Fourier component of velocity onto the displacement vectors, 
and performed inverse Fourier transformation of each decomposed component in real-space. 

\begin{figure}[t]
 \centering
  \includegraphics[width=8.cm,clip]{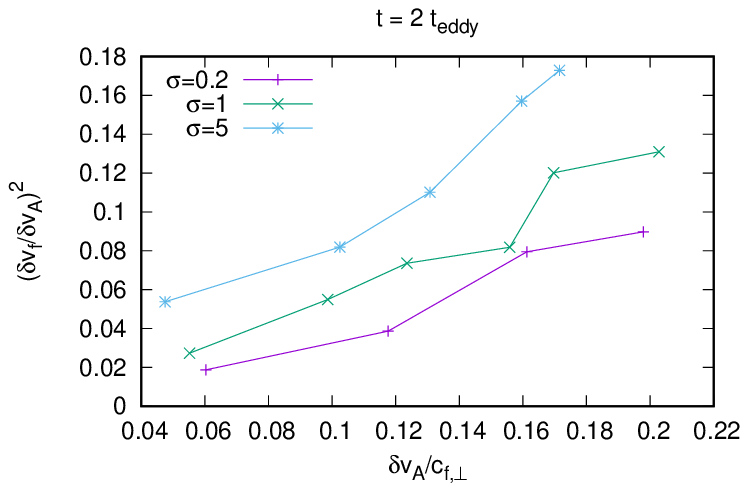}
  \includegraphics[width=8.cm,clip]{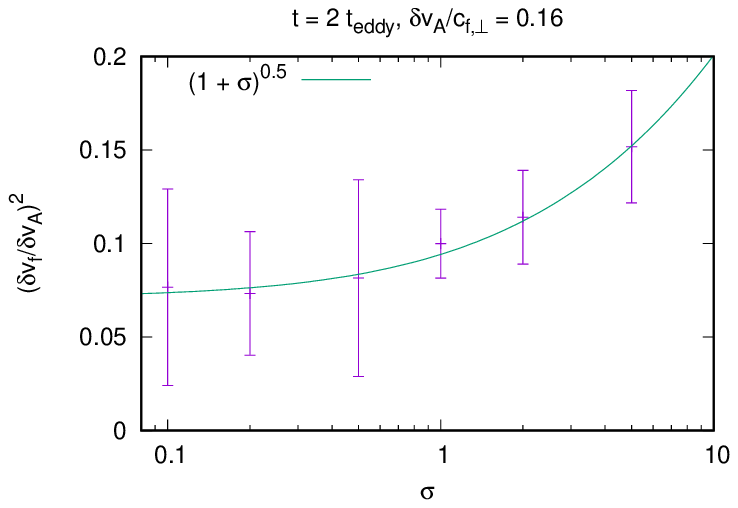}
  \caption{Left: The ratio of fast to Alfv\'en mode velocity power 
           in terms of the non-relativistic fast Mach number %perpendicular to magnetic field 
           of the Alfv\'en mode component using 3-velocity.
%           This indicates Alfv\'en mode saturates around $\delta v_{\rm A}/c_{\rm f,\perp} \sim 0.25$ in the case of 
%           $t = t_{\rm eddy} \equiv L/\delta v_{\rm inj}$. 
           This indicates that the compressible mode becomes more important with increasing $\sigma$-parameter.
           Right: The ratio of fast to Alfv\'en mode velocity power 
           in terms of the background $\sigma$ parameter at $t = 2 t_{\rm eddy}$ and $\delta v_{\rm A}/c_{\rm f,\perp} = 0.16$ 
           whose values are obtained by fitting of the curves of each $\sigma$-value by linear curves. 
           Note that the error bar results from the fitting of the curves. 
           This shows that the ratio is proportional with $\sqrt{1 + \sigma}$ when $\sigma > 1$.}
  \label{fig:1}
\end{figure}

\section{\label{sec:sec3}Results}
The left panel of Figure \ref{fig:1} is the ratio of fast to Alfv\'en mode velocity power at 2 eddy-turnover time 
in terms of the fast Mach number of the Alfv\'en component of velocity. 
Each point corresponds to a simulation result with an injection velocity. % $\delta v_{\rm inj}$. 
Note that the fast velocity $c_{\rm f,\perp}$ in the horizontal axis is taken as that in the perpendicular direction to the background magnetic field. 
%First, we discuss sub-fast regime. 
%In the low-$\sigma$ case, equivalent to the non-relativistic electromagnetic field case, 
The curves in the panel show that 
the fast mode power increases nearly linearly with the fast Mach number %irrespective of their $\sigma$-parameter, 
as reported in the non-relativistic case \citep{2002PhRvL..88x5001C,2003MNRAS.345..325C}. 
%On the other hand, it is clear that the fast component increases with $\sigma$-parameter 
In addition, 
it also shows that the fast component increases with $\sigma$-value.  
%when $\sigma$ becomes larger than unity, the Poynting-dominated regime. 
The right panel of Figure \ref{fig:1} is the ratio of fast to Alfv\'en mode velocity power in terms of the background $\sigma$ value 
at $t = 2 t_{\rm eddy}$ and $\delta v_{\rm A}/c_{\rm f,\perp} = 0.16$. 
It shows the ratio is nearly independent of the $\sigma$-parameter when $\sigma < 1$ 
whose value is around 0.08. 
Note that this indicates 
the Alfv\'en to fast energy conversion is not more than $8 \%$ in the low-$\sigma$ plasma, 
which is consistent with the result obtained by \citep{2002PhRvL..88x5001C}. 
On the other hand, it increases approximately by $(1 + \sigma)^{1/2}$ when $\sigma > 1$, 
which is basically consistent with our previous result of the driven turbulence \citep{2015ApJ...815...16T}. 
This indicates that the ratio can be written as: 
\begin{align}
  (\delta v_{\rm f})^2 / (\delta v_{\rm A})^2 &\propto  (\delta v_{\rm A})/c_{\rm fast,\perp} \quad & ({\rm when} \ \sigma \ll 1)
  ,
  \label{eq:4.1}
  \\                                              
                                            &\propto (1 + \sigma)^{1/2} (\delta v_{\rm A})/c_{\rm fast,\perp} \quad & ({\rm when} \ \sigma \gtrsim 1)
  \label{eq:4.2}
  .
\end{align}
Note that this is a relativistic extension of the Equation (6) in \citep{2002PhRvL..88x5001C}. 
%However, the left panel of Figure \ref{fig:1} shows that 
%this equation is valid only in the sub-fast regime. 
%When the fast Mach number of the Alfv\'en mode velocity is comparable to $0.25$, 
%the linear growth of the ratio stops and the ratio starts to increase its value at a nearly fixed value of $\delta v_{\rm A}/c_{\rm f,\perp}$. 
This indicates that the 
non-linearity of the Alfv\'en mode becomes strong enough to convert its energy into compressible mode at this value of fast Mach number 
as the electromagnetic field becomes relativistically strong. 
%\footnote{
%As a result of the relativistic magnetic field, 
%the electric field, not taken into account in the non-relativistic case, becomes important in this regime, 
%which drives compressible modes more efficiently when $\sigma > 1$. 
%More detailed explanation will be discussed in our forthcoming papers. 
%}. 
%Actually, 
%the data points deviating from Equations (\ref{eq:4.1}) and (\ref{eq:4.2}) corresponding to the cases whose injection velocity is larger than $0.7 c_A$, 
%that is, trans-Alfv\'enic cases. 
%Note that left-panel of Figure \ref{fig:1} indicates that 
%the fast to Alfv\'en mode power ratio saturates around $(\delta V_{\rm fast}/ \delta V_{\rm Alfven})^2 \lesssim 0.2$ for all $\sigma$ value 
%even in the strongly non-linear regime where the curves do not follow Equations (\ref{eq:4.1}) and (\ref{eq:4.2}) 
%because of the fast shock conversion and dissipation.
%Note that non-linearity becomes strong at much smaller value of the turbulence fast Mach number $\delta v_{\rm A}/c_{\rm f, \perp}$ 
%than the non-relativistic case reported by \citep{2002PhRvL..88x5001C,2003MNRAS.345..325C} 
%that found the linear relation, Equation (\ref{eq:4.1}), continues until around $\delta v_{\rm A}/c_{\rm A} \sim 0.7$. 
%\textcolor{red}{%
Qualitatively, 
this can be explained by the electric field that should be taken into account in the relativistic MHD equations %(\ref{eq:1.1}) - (\ref{eq:1.3}) 
because of the relativistic velocity: $|{\bf E}| = |- ({\bf v}/c) \times {\bf B}| \sim |{\bf B}|$. 
If we use the quasi-linear theory, that is, take into account the 2nd-order terms in equations assuming 1st-order Alfv\'en mode perturbation, 
%the force term in the equation of motion of RMHD, Equation (\ref{eq:1.2}), can be written as: 
the force term in the equation of motion of RMHD can be written as: 
\begin{align}
  F_{\rm x} &\equiv - \nabla_x \left[ \frac{B_0^2}{2} \left( 1 + \frac{c_{\rm A}^2}{c^2} \right) \left(\frac{\delta v_{\rm A}}{c_{\rm A}}\right)^2 \right]
  ,
  \\            
  F_{\rm y} &\equiv - \nabla_y \left[ \frac{B_0^2}{2} \left( 1 - \frac{c_{\rm A}^2}{c^2} \right) \left(\frac{\delta v_{\rm A}}{c_{\rm A}}\right)^2 \right]
  ,
  \\
  F_{\rm z} &\equiv - B_0 \nabla_x \delta B_{\rm A}
  ,
\end{align}
where we set the background magnetic field as ${\bf B}_0 = B_0 {\bf e}_x$ and the Alfv\'en mode direction in z-direction. 
The z-component drives the usual Alfv\'en mode. 
Interestingly, 
the anisotropic nature of the electric field gives the weaker force in the  y-direction and the stronger force in x-direction, 
or the parallel direction of the background magnetic field. 
In the Poynting-dominated regime, $c_{\rm A} \sim c$, 
the force is only in the direction parallel to the magnetic field direction, 
and we consider this is the origin of increasing fast mode conversion. 
%}%
Note that this also indicates that the mode coupling between fast and Alfv\'en modes becomes stronger, 
which is not observed in non-relativistic MHD turbulence \citep{2002PhRvL..88x5001C}. 
%\textcolor{red}{%
Rewriting Equation (\ref{eq:4.2}) as, $(\delta v_{\rm F}/\delta v_{\rm A})^2 \simeq \alpha \sqrt{1 + \sigma}$, 
where $\alpha \equiv 0.4 \delta v_{\rm A}/c_{\rm fast,\perp}$ 
whose coefficient $0.4$ is obtained by the fitting data in Figure \ref{fig:1}, 
the completely coupling of fast and Alfv\'en modes, $\delta v_{\rm F} \sim \delta v_{\rm A}$, occurs
when $\sigma \sim 1/\alpha^2 - 1$; 
If we assume $\delta v_{\rm A}/c_{\rm fast,\perp} \sim 0.5$, 
the necessary $\sigma$ value becomes around 24. 
In this regime, we can expect an appearance of a new RMHD turbulence regime 
where the critical balance is invalid due to the strong coupling to the fast mode. 
%}%
Concerning the slow mode conversion, 
we found that 
slow modes also show similar behavior to the fast mode as shown in Figure \ref{fig:1}, 
but their kinetic energy is approximately 1.5 times larger than the fast mode case. 
We take this to be due to the fact that 
the considered turbulence is in the regime of sub-Alfv\'enic but super-slow turbulence. 
%We consider 
%this is because the considered turbulence is in the regime of sub-Alfv\'enic but super-slow turbulence. 
%More detailed analysis will be reported in our successive papers. 
More detailed analysis will be reported in our forthcoming papers. 

Importantly, 
the fast to Alfv\'en mode power ratio depends on a 3-fast Mach number, not the relativistic 4-fast Mach number. 
This means that in Poynting-dominated plasma the compressible mode turbulence becomes important 
even if the kinetic energy of the turbulence is much smaller than the background magnetic field energy, 
and this is very different from the non-relativistic case whose kinetic energy of turbulence should be comparable to the background magnetic field energy, 
$\rho v_{\rm turb}^2 \sim B^2$. 
This will be important for the electron and cosmic-ray acceleration by MHD turbulence \citep{1949PhRv...75.1169F,1954ApJ...119....1F} in high-energy astrophysical phenomena, 
such as GRBs and blazars (see e.g. \citep{2015ApJ...808L..18A}) 
%We will discuss this in our succeeding papers. 
\footnote{
%\textcolor{red}{%
The enhancement of compressibility effects was mentioned in \citep{2013ApJ...766L..10R} 
in the relativistic temperature case, but in the paper we quantify this effects 
covering Poynting-energy dominated regime. 
%}%
}

\begin{figure*}[t]
 \centering
  \includegraphics[width=16.cm,clip]{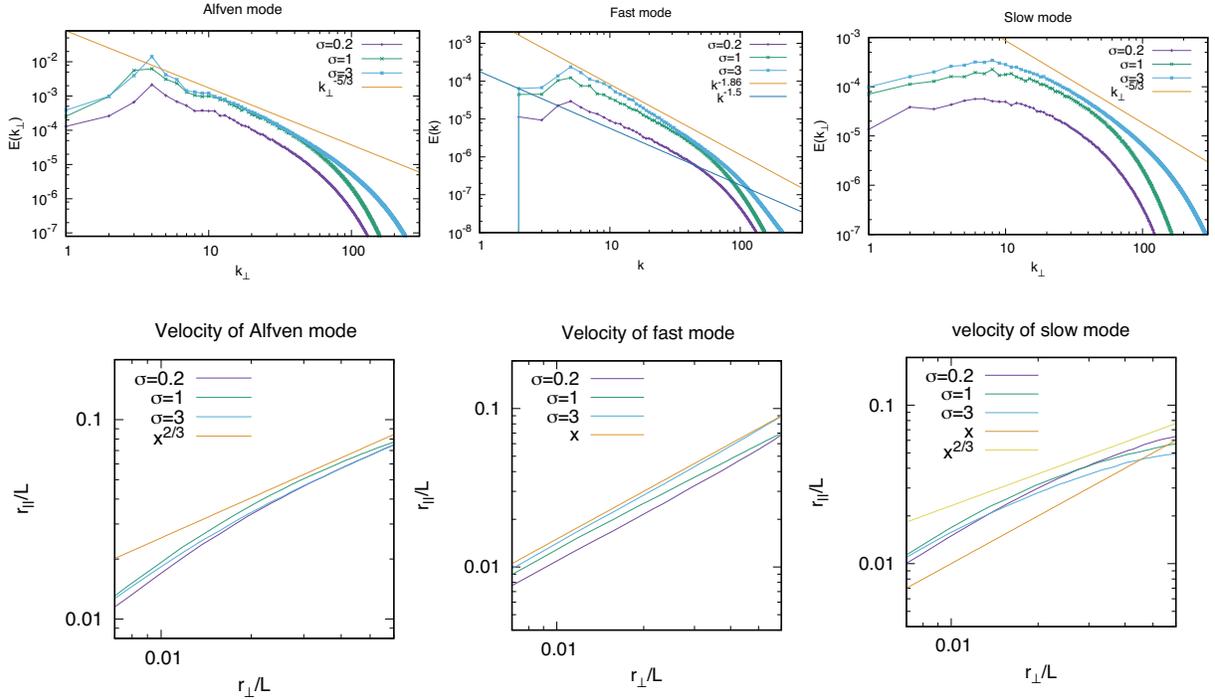}
  \caption{Top: Kinetic energy spectra of Alfv\'en, fast, and slow mode. 
           Bottom: Eddy-scale of Alfv\'en, fast, and slow mode obtained by a 2nd-order velocity structure function.
           All the data are measured at 1 eddy-turnover time. 
           The initial Alfv\'en mode turbulence is injected at $k/2 \pi = 3/L$ with velocity dispersion 
           $\delta v/c_{\rm A} = 0.6$ for $\sigma = 0.2, 1$ and $\delta v/c_{\rm A} = 0.5$ for $\sigma = 3$. 
           }
  \label{fig:2}
\end{figure*}

\textit{Alfv\'en Mode}--The top-left panel of Figure \ref{fig:2} shows the kinetic energy spectrum of Alfv\'en mode 
in terms of the wave vector perpendicular to the background magnetic field. 
%It shows the spectrum follows the Kolmogorov spectrum in its inertia region: 
It indicates that the spectrum follows the Kolmogorov spectrum in its inertial region: 
\begin{equation}
  \label{eq:5.1}
  E^{\rm A} (k_{\perp}) \propto k_{\perp}^{-5/3}
  ,
\end{equation}
%as reported by \citep{2012ApJ...744...32Z,2013ApJ...763L..12Z,2013ApJ...766L..10R} 
%that performed not the mode decomposition but the Helmholtz decomposition of velocity 
%differently from this Letter. 
which is consistent with the critical balance predicted by \citep{1995ApJ...438..763G}, 
%\textcolor{red}{%
meaning the energy-cascade time by Alfv\'en mode along the magnetic field is comparable to that by eddy-interaction perpendicular to the magnetic field, 
and the turbulent eddy becomes anisotropic
%}%
%in particular, the low-$\sigma$ case, or the matter-dominated case
\footnote{
Note that it is still very difficult to judge if the spectrum index is $-5/3$ or $-1.5$ due to the insufficient numerical resolution. 
%\textcolor{red}{%
The effects of the bottleneck \citep{2010ApJ...722L.110B,2014ApJ...784L..20B,2015ASSL..407..163B} might also distort the measured spectra. 
%}%
}. 
Note that this is also consistent with the results by \citep{2012ApJ...744...32Z,2013ApJ...763L..12Z,2013ApJ...766L..10R} 
that performed not the mode decomposition but the Helmholtz decomposition of velocity, 
although the incompressible mode obtained by the Helmholtz decomposition is essentially different from the Alfv\'en mode. 
The bottom-left panel of Figure \ref{fig:2} shows the values of $r_{||}$ and $r_{\perp}$ with the same 2nd-order structure function for the velocity. 
The distance $r_{||}$ and $r_{\perp}$ is determined using the local magnetic field direction 
following \citep{2000ApJ...539..273C}.
%\textcolor{red}{%
Note that it is essential to measure $r_{||}$ and $r_{\perp}$ in terms of the local mean magnetic field 
because the critical balance proposed by \citep{1995ApJ...438..763G} is applicable only in the local system of reference, 
which was established by \citep{1999ApJ...517..700L,2000ApJ...539..273C,2001ApJ...554.1175M,2002ApJ...564..291C}. 
%}%
It shows that the eddy size scales as $k_{||} \propto k_{\perp}^{2/3}$ 
as predicted by \citep{1995ApJ...438..763G}. 
Our simulation results indicate that 
%the critical balance is still valid in high-$\sigma$ regime (as reported in Cho \& Lazarian 2014??). 
the critical balance is still valid in high-$\sigma$ regime 
which is consistent with works assuming the force-free approximation \citep{1998PhRvD..57.3219T,2005ApJ...621..324C,2014ApJ...780...30C}
\footnote{
Note that we observed only the \textit{strong turbulence} regime, though we injected sub-Alfv\'enic turbulence 
that induces the \textit{weak turbulence} in the large scale regime as first pointed out in \citep{1999ApJ...517..700L}. 
We take this to indicate that the resolution in the direction perpendicular to the background magnetic field is not enough to observe the weak turbulence cascade. 
It may also be because the generation of compressible modes hinders the Alfv\'en mode energy in cascading in the perpendicular direction 
(pointed out by S\'ebastien Galtier). 
Recent findings of the weak turbulence regime in numerical simulations are reported by \citep{2016PhRvL.116j5002M}.
}
.

\textit{Fast Mode}--The top-middle panel of Figure \ref{fig:2} shows the kinetic energy spectrum of fast mode 
in terms of the wave vector perpendicular to the background magnetic field. 
It indicates that the energy spectrum of their inertial region can be written as: 
\begin{align}
  E^{\rm f} (k) &\propto k^{-3/2}  \quad & ({\rm when} \ \sigma < 1),
  \label{eq:5.2}
  \\
  &\propto k^{-1.86}  \quad & ({\rm when} \ \sigma > 1)
  \label{eq:5.2.2}
  .
\end{align}
%Note that the spectrum index $-1.86$ is consistent with the previously obtained value by \citep{2012ApJ...744...32Z,2013ApJ...763L..12Z} 
%which did not perform mode decomposition but considered the compressible component (potential component) ${\bf u}_{\rm p} = \nabla \phi$ 
%where $\phi$ is a scalar function. 
%\textcolor{red}{%
The value of the spectrum index $1.86$ is previously obtained by \citep{2012ApJ...744...32Z,2013ApJ...763L..12Z} 
which did not perform mode decomposition but considered the compressible component (potential component) ${\bf u}_{\rm p} = \nabla \phi$ 
where $\phi$ is a scalar function. 
Note that our results indicates that the index becomes slightly larger than $1.86$, and an increasing function of $\sigma$. 
%}%
The bottom-middle panel of Figure \ref{fig:2} shows the values of $r_{||}$ and $r_{\perp}$ of fast mode turbulent eddy. 
It indicates that the eddy is nearly isotropic, $r_{||} \propto r_{\perp}$ independent of its scale, 
similar to the non-relativistic case. 

\textit{Slow Mode}--The top-right panel of Figure \ref{fig:2} shows the kinetic energy spectrum of slow mode 
in terms of the wave vector perpendicular to the background magnetic field. 
%The indicated energy spectrum seems: 
%The indicated energy spectrum can be written as; 
%\begin{equation}
%  E^{\rm s} (k_{\perp}) \propto k_{\perp}^{-5/3}
%  \label{eq:5.3}
%  .
%\end{equation}
%\textcolor{red}{%
%The indicated energy spectrum seems difficult to be fitted by power laws 
%whose form do not change even if the resolution is increased from $L/512$ to $L/2024$. 
%We consider this indicates that 
%the coupling between slow mode and Alfv\'en becomes stronger, and 
%the slow mode energy does not cascade but change into the other modes. 
%However, the too small size of the inertial region makes it unclear if it is correct or not. 
%The indicated energy spectrum seems difficult to be fitted by power laws 
%whose form do not change even if the resolution is increased from $L/512$ to $L/2024$. 
%Therefore we do not make strong statements about the slow mode slope, but note that 
%the measured anisotropies agree better than the slope with the theoretical predictions.
The indicated energy spectrum indicates the non-power law behavior even as the resolution
is increased from L/512 to L/2024. This may signify that the energy exchange between
slow mode and Alfven becomes stronger than in the case of non-relativistic turbulence in \citep{2002PhRvL..88x5001C}. 
%}%
The bottom-right panel of Figure \ref{fig:2} shows the values of $r_{||}$ and $r_{\perp}$ of the slow mode turbulent eddy. 
It indicates that the slow mode eddy size also follows the critical condition, similar to the Alfv\'en mode. 
We take these results to indicate that 
the slow mode, or the pseudo-Alfv\'en mode, do not cascade its energy for themselves 
as is known to occur in the non-relativistic case \citep{2001ApJ...562..279L,2002PhRvL..88x5001C}.

\section{\label{sec:sec4}Discussion and Conclusion}
In this Letter, we investigated the properties of each characteristic mode of the relativistic ideal MHD turbulence. 
%\textcolor{red}{%
The most important finding is that 
the ratio of fast to Alfv\'en mode velocity power increases not only with the fast Mach number of the Alfv\'en mode velocity, 
but with the background $\sigma$-parameter when $\sigma > 1$. 
This indicates that the mode coupling between fast and Alfv\'en modes becomes stronger in Poynting-dominated plasmas, 
which is not observed in non-relativistic MHD turbulence. 
This also suggests that 
a new turbulence regime will appear in a sufficiently high-$\sigma$ plasma 
where Alfv\'en mode and fast mode completely couples, and the critical-balance cannot be applied. 
%}%
We also investigated the behavior of the energy spectra and eddy shapes. 
Although Alfv\'en and slow modes show a similar behavior to their non-relativistic cases,  
the energy spectrum of fast mode shows a different behavior 
whose spectrum index increases from $3/2$ to more than $1.86$ with $\sigma$-parameter. 
Our finding will advance our understanding of high-energy astrophysical phenomena, 
for example, 
%the high-energy photon radiation from various high energy astrophysical phenomena 
%whose electrons are considered to be accelerated by strong MHD turbulence. 
the acceleration of electrons resulting in the observed non-thermal photon energy spectrum \citep{1949PhRv...75.1169F,1954ApJ...119....1F,2004ApJ...610..550P,2014ApJ...784...38L}, 
enhancement of magnetic reconnection responsible for relativistic jet acceleration \citep{2015ApJ...815...16T,2016arXiv160303276S}, 
and cosmic-ray diffusion in turbulent media \citep{2014ApJ...784...38L}. 
It also indicates that 
the compressible turbulence will be involved in many high-energy astrophysical objects with Poynting-dominated plasma, 
such as the pulsar wind nebulae, relativistic jets, and GRBs. 

\acknowledgments
We would like to thank S\'ebastien Galtier, Supratik Banerjee, and Nobumitsu Yokoi 
for many fruitful comments and discussions. 
%We also would like to thank our anonymous referee for a lot of fruitful comments on our paper. 
Numerical computations were carried out on the Cray XC30 
at Center for Computational Astrophysics, CfCA, of National Astronomical Observatory of Japan.
Calculations were also carried out on SR16000 at YITP in Kyoto University. 
This work is supported in part by the Postdoctoral Fellowships by the Japan Society for the Promotion of Science No. 201506571 (M. T.). 
%One of the auther (A. Lazarian) is supported by NSF AST 1212096. 
%\textcolor{red}{%
AL is supported by NSF AST 1212096, NASA grant X5166204101. % and of the NSF sponsored Center for Magnetic Self-Organization. 
%}%

%SR16000'à

%% To help institutions obtain information on the effectiveness of their
%% telescopes, the AAS Journals has created a group of keywords for telescope
%% facilities. A common set of keywords will make these types of searches
%% significantly easier and more accurate. In addition, they will also be
%% useful in linking papers together which utilize the same telescopes
%% within the framework of the National Virtual Observatory.
%% See the AASTeX Web site at http://www.journals.uchicago.edu/AAS/AASTeX
%% for information on obtaining the facility keywords.

%% After the acknowledgments section, use the following syntax and the
%% \facility{} macro to list the keywords of facilities used in the research
%% for the paper.  Each keyword will be checked against the master list during
%% copy editing.  Individual instruments or configurations can be provided 
%% in parentheses, after the keyword, but they will not be verified.

%%{\it Facilities:} \facility{Nickel}, \facility{HST (STIS)}, \facility{CXO (ASIS)}.

%% Appendix material should be preceded with a single \appendix command.
%% There should be a \section command for each appendix. Mark appendix
%% subsections with the same markup you use in the main body of the paper.

%% Each Appendix (indicated with \section) will be lettered A, B, C, etc.
%% The equation counter will reset when it encounters the \appendix
%% command and will number appendix equations (A1), (A2), etc.

%\bibliography{RCT}% Produces the bibliography via BibTeX.

\end{document}